\documentclass[aps,pre,preprint,showpacs]{revtex4}
\usepackage{amsmath,amssymb}

\begin{document}

\title{Modified kinetic theory of Bose systems taking into account
 slow hydrodynamical processes}

\author{P.A.~Hlushak}\email[]{phl@icmp.lviv.ua}
\author{M.V. Tokarchuk}
\affiliation{Institute for Condensed Matter Physics  of the National
Academy of Sciences of Ukraine,\\ 1~Svientsitskii~Str., UA--79011 Lviv,
Ukraine}

\begin{abstract}

An approach to the description of  kinetics which taking into account the
large-scale hydrodynamic transport processes for quantum Bose system is
proposed. The nonequilibrium statistical operator  which consistently
describes both the kinetic and nonlinear hydrodynamic fluctuations in
quantum liquid is calculated. Using this operator the coupled equations
for quantum one-particle distribution function and functional of
hydrodynamic variables: densities of momentum, energies and number of
particles are obtained.
\end{abstract}


\maketitle

\section{Introduction}

The examination of dynamic properties of quantum liquids as well as of the
features of transition processes from gaseous state to fluid and
superfluid one with temperature decrease  remains a hard problem in modern
physics. The development of the nonequilibrium statistical theory which
would take into account one-particle and collective physical processes
that occur in a system is an example of such problem. The problem the
going out beyond the hydrodynamic area to the area of an intermediate
values of wave vector and frequency, where the kinetic and hydrodynamic
processes are interdependent and should be considered simultaneously, is
essential. The separation of contributions from the kinetic and
hydrodynamic fluctuations into time correlation functions, excitations
spectrum, transport coefficients allows one to obtain much more
information on physical processes with the different time and spatial
intervals, which define the dynamic properties of the system.

The quantum system of Bose particles serves as a physical model in
theoretical descriptions both the equilibrium and nonequilibrium
properties of real helium. In particular, many articles
\cite{l1,l2,l3,l4,l5,l6,l7,l8,l9,l10,l11,l12,l13,l14,l15,l16,l17,l18,l19,l20,l21,l22}
are devoted to the hydrodynamic description of normal and superfluid
states of such system. A brief review of the results of the investigations
within the framework of linear hydrodynamics has been considered in the
article by Tserkovnikov \cite {l20}. In papers \cite{l23,l24,l25}
theoretical approaches are proposed to the description of nonlinear
hydrodynamic fluctuations connected with problem of calculating the
dispersion for the kinetic transport coefficients and the spectrum of
collective modes in the low-frequency area for superfluid Bose liquid. The
generalized Fokker-Planck equation for the nonequilibrium distribution
function of slow variables for quantum systems was obtained in paper by
Morozov \cite{l26}. Problems of building the kinetic equation for Bose
systems based on the  microscopic approach were considered in papers
\cite{l26,l28}. For normal Bose systems, the calculations of the
collective mode spectrum (without accounting for a thermal mode), dynamic
structure factor, kinetic transport coefficients \cite[see the
references]{l13} are carried out on the basis of the hydrodynamic or
kinetic approaches. Nevertheless, these results are valid only in the
hydrodynamic area (small values of wave vector $\bf k$ and frequency
$\omega$). The papers \cite {l28, l29,l30} were devoted to the
investigation of the dynamic structure factor and of collective
excitations spectrum for superfluid helium.

In papers \cite {l31,l32}, a generalized scheme for the theoretical
description of dynamic properties of semiquantum helium has been proposed
based on the method of nonequilibrium statistical operator. Here the set
of equations of the generalized hydrodynamics is obtained and the thermal
viscous model with kinetic and hydrodynamical collective modes is analyzed
in details. The closed system of the equations for time correlation
functions is obtained within the Markovian approximation for transport
kernels. Using these equations the analysis of dynamic properties of
semiquantum helium is carried out at two values of temperature above
transition to a superfluid state. Similar investigations were performed in
papers \cite {l34,l35,l36} for helium above a point of the phase
transition.

In general,  a hard problem exists in  the description of Bose systems
going out from the hydrodynamic area to the area of intermediate values of
$ \bf k $ and $ \omega $, where the kinetic and hydrodynamic processes are
interdependent and should be considered simultaneously. This is one of the
urgent problems of the statistical theory of nonequilibrium transport
processes in quantum liquid. It should be noted that in the paper by
Tserkovnikov \cite {l36}, a problem of building of the linearized kinetic
equation for the Bose system above critical temperature was considered by
means of the method of two-time Green functions \cite{l37,l38}.

The investigations of semiquantum helium \cite{l31,l32} became a main step
in this direction. A considerable success was achieved in papers \cite
{l40,l41,l42} in which the approach of the consistent description of
kinetics and hydrodynamics of classical dense gases and fluids is proposed
based on the method of Zubarev nonequilibrium statistical Zubarev operator
\cite{l23,l44,l45,l47}. By means of this formalism the nonequilibrium
statistical operator of many-particles Bose system, which consistently
describes kinetics and hydrodynamics, is obtained in papers
\cite{l48,l49}. The quantum nonequilibrium one-particle distribution
function and the average value of density of interaction potential energy
have been selected as parameters of consistent description of a
nonequilibrium state.

On the other hand, the large-scale fluctuations in a system related to the
slow hydrodynamical processes play the essential role at the phase
transition with temperature decrease. The construction of the kinetic
equations taking into account the slow processes is the hard problem for
the transport theory both in classical and quantum liquids. The same
problem arises at the description of low-frequency anomalies in kinetic
equations related to "long tails" of correlation functions
\cite{l50,l51,l52,l53,l54} as well as at the consistent description of
collective effects in plasma \cite{l55}.

The aim of the present paper is construction of the kinetic equations for
quantum system taking into account the nonlinear hydrodynamic processes
using the nonequilibrium statistical operator method.

In the second part of the paper we shall obtain the nonequilibrium
statistical operator and generalized transport equations for quantum Bose
system, when the nonequilibrium single-particle Wigner distribution
function and nonequilibrium average operator of potential interaction
energy are chosen as basic parameters of the abbreviated description.

In the third part the approach to the description of kinetics taking into
account the slow hydrodynamical transport processes for quantum Bose
system is considered. The nonequilibrium statistical operator which
consistently describes both the kinetic and nonlinear hydrodynamical
fluctuations in a quantum liquid is calculated. The coupled set of kinetic
equations for quantum one-particle distribution function and generalized
Fokker-Plank equations  for the functional of hydrodynamical variables:
particles number, momentum and energy densities  is obtained. Neglecting
the hydrodynamic fluctuations, we obtain the traditional scheme of kinetic
theory.

\section{The nonequilibrium statistical operator of the consistent
 description of kinetics and
 hydrodynamics of quantum Bose system}

Observable average values of energy density
$\langle\hat{\varepsilon}_{\bf{q}}\rangle^t$,  momentum density
$\langle\hat{P}_{\bf{q}}\rangle^t$, and particle numbers density
$\langle\hat{n}_{\bf{q}}\rangle^t$ are the abbreviated description
parameters at investigations of the hydrodynamical nonequilibrium state of
the normal Bose liquid which is characterized by processes of the energy,
momentum and masses flows. Operators for these physical quantities are
defined through the Klimontovich operator of the phase particle number
density $\hat{n}_{\bf q}({\bf p}) = \hat {a}^{+}_{\bf{p - \frac{q}{2}}}
\hat {a}_{\bf{p+\frac{q}{2}}} $:
\begin{equation}
\label{eq2.1}
\hat {n}_{\bf q} = \frac{1}{\sqrt N }\sum\limits_{\bf{p}}
{\hat {n}_{\bf q} ({\bf p})},
\end{equation}
\begin{equation}
\label{eq2.2}
\hat {\bf P}_{\bf q} =
\frac{1}{\sqrt N }\sum\limits_{\bf p} {{\bf p}\,\hat{n}_{\bf q} ({\bf p})},
\end{equation}
\begin{equation}
\label{eq2.3}
\hat {\varepsilon }_{\bf q}^{kin}  = \frac{1}{\sqrt N }\sum\limits_{\bf p}
{\left( {\frac{p^2}{2m} - \frac{q^2}{8m}} \right)\hat {n}_{\bf q} ({\bf p})},
\end{equation}
\begin{equation}
\label{eq2.4}
\hat {\varepsilon }_{\bf q}^{int} = \frac{1}{\sqrt N }\sum\limits_{\bf p}
\sum\limits_{{\bf p}'}{\sum\limits_{\bf k} \nu (k) \hat {a}^{+}_{{\bf p} +
\frac{{\bf k} - {\bf q}}{2}} \, \hat {n}_{\bf q} ({\bf p}')
\hat {a}_{{\bf p} - \frac{{\bf k} - {\bf q}}{2}}}   ,
\end{equation}
 where
$\hat{\varepsilon}_{\bf q}^{kin}$ and $\hat{\varepsilon}_{\bf{q}}^{int}$
are Fourier-components of the operators of kinetic and potential energy
densities. Average value of the phase particles number density operator is
equal to the nonequilibrium one-particle distribution function
$f_1({\bf{q,p}},t)=\langle\hat{n}_{\bf{q}}({\bf p})\rangle^t$, which
satisfies the kinetic equation for quantum Bose system.

The agreement between kinetics and hydrodynamics  for dilute Bose gas does
not cause problems because in this case the density is a small parameter.
Therefore, only  the quantum one-particle distribution function
$f_1({\bf{q}}, {\bf{p}};t)$ can be chosen for parameter of the abbreviated
description. At transition to quantum Bose liquids, the contribution of
collective correlations, which are described by average potential energy
of interaction,  is more important than one-particle correlations
connected with $f_1 ({\bf q}, {\bf p}; t) $. From this fact it follows
that for consistent description of kinetics and hydrodynamics of Bose
liquid, the one-particle nonequilibrium distribution function along with
the average potential energy of interaction are necessarily should be
chosen as the parameters of the abbreviated description \cite{l48,l49}.
The nonequilibrium state of such quantum system is completely described by
the nonequilibrium statistical operator $\hat{\varrho}(t)$ which satisfies
the quantum Liouville equation:
\begin{equation}
\label{eq2.5}
 \frac{\partial}{\partial t}\hat{\varrho}(t)+ i\hat{L}_N \hat{\varrho}(t) =  -
\varepsilon \left( \hat {\varrho }(t) - \hat {\varrho }_q
(t)\right).
\end{equation}
The infinitesimal source $\varepsilon$ in the right-hand side of this
equation breaks symmetry of Liouville equation with respect to $t\to\,-t$
and selects retarded solutions ($\varepsilon\to\,+0$ after limiting
thermodynamic transition). The quasiequilibrium  statistical operator
$\hat{\varrho}_q(t)$ is determined from the condition of the informational
entropy extremum of systems at the conservation of normalization condition
$\mbox{Sp}\,\hat{\varrho}_q(t)=1$ for fixed values $\langle
\hat{n}_{\bf{q}} ({\bf{p}})\rangle^t$ and
$\langle\hat{\varepsilon}^{int}_{\bf{q}} \rangle^t$ \cite{l48,l49}:
\begin{equation}
\label{eq2.6}
\hat{\varrho }_q (t) = \exp
\left\{
  -\Phi(t)- \sum\limits_{\bf q}
  \beta _{-{\bf q}}(t) \hat{\varepsilon}^{int}_{\bf q}  -
  \sum\limits_{\bf q} \sum\limits_{\bf p} \gamma _{- {\bf q}} ({\bf p};t)
  \hat {n}_{\bf q} ({\bf p})
\right\},
\end{equation}
\noindent
where the Lagrangian multipliers $\beta_{-{\bf q}}(t)$,
$\gamma_{-{\bf{q}}}({\bf{p}};t)$ are determined from the self-consistent
conditions:
$$
\langle \hat{n}_{\bf q}({\bf p})\rangle^t=\langle \hat{n}_{\bf{q}}({\bf p})\rangle^t_{q},
\quad
\langle \hat {\varepsilon }^{int}_{\bf q}\rangle^t=
\langle\hat{\varepsilon}^{int}_{\bf q}\rangle^t_{q}.
$$
The Massieu-Plank functional
\begin{equation}
\label{eq2.7}
\Phi (t) = \ln \,\mbox{Sp}\,\exp \left\{- \sum\limits_{\bf q}
  \beta _{-{\bf q}}(t) \hat{\varepsilon}^{int}_{\bf q}  -
  \sum\limits_{\bf q} \sum\limits_{\bf p} \gamma _{- {\bf q}} ({\bf p};t)
  \hat {n}_{\bf q} ({\bf p})
\right\}
\end{equation}
is determined from the normalization condition. Here
$\langle(...)\rangle^t=\mbox{Sp}(...)\hat{\varrho}(t)$,
$\langle(...)\rangle^t_{q}=\mbox{Sp}(...)\hat{\varrho}_{q}(t)$. At given
quasiequilibrium statistical operator $\hat{\varrho}_q(t)$ we can find the
nonequilibrium statistical operator $\hat{\varrho}(t)$ that satisfies the
quantum Liouville equation in the presence of a source:
\begin{eqnarray}
\label{eq2.8}
 \hat{\varrho}(t) =\hat{\varrho}_q (t)+
 \sum\limits_{\bf q} \int\limits_{-\infty}^t dt'
 \mbox{e}^{\varepsilon (t'-t)} T_q (t,t')
\int\limits_0^1 {d\tau } \left( \hat{\varrho}_q (t) \right)^\tau
 I_\varepsilon^{int}({\bf q},t')
 \left( {\hat{\varrho}_q (t)} \right)^{1 - \tau }\beta_{-{\bf q}}(t')
 \\
 + \sum\limits_{\bf q} \sum\limits_{\bf p} \int\limits_{-\infty }^t dt'
 \mbox{e}^{\varepsilon (t' - t)} T_q (t,t')
\times \int\limits_0^1 {d\tau } \left( \hat{\varrho}_q (t) \right)^\tau I_n({\bf p},{\bf q},t')
 \left( \hat {\varrho }_q (t)\right)^{1 - \tau }\gamma_{-{\bf q}} ({\bf p},t'),
\nonumber
\end{eqnarray}
where the generalized flows
\begin{equation} \label{eq2.9}
I_{\varepsilon}^{int}({\bf q},t)=\Big( 1-P(t) \Big) i\hat L_N\hat\varepsilon_{\bf q}^{int},
\end{equation}
\begin{equation} \label{eq2.10}
I_n({\bf p},{\bf q},t)=\Big( 1-P(t) \Big) i\hat L_N \hat n_{\bf q}(\bf p)
\end{equation}
contain generalized Mori projection operator. $T_q (t, t ')$ is the
generalized evolution Kawasaki-Gunton operator with the projection
\cite{l48,l49}. The nonequilibrium statistical operator (\ref{eq2.8}) is
obtained at abbreviated description of kinetics and hydrodynamics of Bose
system. Using it, we can find the non-closed system of transport equations
for the parameters  of  abbreviated description
$\hat{n}_{\bf{q}}({\bf{p}})\rangle^t$ and
$\langle\hat{\varepsilon}_{\bf{q}}^{int}\rangle^t$ \cite{l5,l6}:
\begin{eqnarray}
\label{eq2.11}
 \frac{\partial}{\partial t}\langle {\hat n}_{\bf q}({\bf p})\rangle^t =
 \langle\dot {\hat n}_{\bf q}({\bf p})\rangle _q^t
+\sum_{\bf q'} \int\limits_{-\infty}^{t}dt'\, \mbox{e}^{ \epsilon
(t'-t)}
 \varphi_{n \varepsilon}^{int}({\bf q},{\bf p},{\bf q'},t,t')
\beta _{-\bf q}(t')
\\
+ \sum_{\bf q'} \sum_{\bf p'} \int\limits_{-\infty}^{t}dt'\,
\mbox{e}^{\epsilon (t'-t)} \, \varphi _{n n}({\bf q},{\bf p},{\bf q'},{\bf q'},t,t')
\gamma _{-\bf q}{(\bf p',t')},
\quad
 \nonumber \\
\frac{\partial}{\partial t}\langle\hat\varepsilon_{\bf q}^{int}\rangle^t=
\langle\dot{\hat \varepsilon}_{\bf q}^{int}\rangle_q^t + \sum_{\bf q'}
\int\limits_{-\infty}^{t}dt'\,\mbox{e}{\epsilon (t'-t)}
\varphi_{\varepsilon \varepsilon}^{int}({\bf q},{\bf q'},t,t') \beta _{-\bf q}(t')
\nonumber \\
+ \sum_{\bf q'} \sum_{\bf p'} \int\limits_{-\infty}^{t}dt'\, \mbox{e}^{\epsilon (t'-t)} \,
 \varphi _{\varepsilon n}^{int,int}({\bf q},{\bf q'},{\bf p'},t,t') \gamma _{-\bf q}({\bf
 p'},t').
\qquad
\nonumber
\end{eqnarray}
In the  equations (\ref{eq2.11})  the generalized transport kernels, which
describe dissipative processes in the system, are introduced as follows:
\begin{equation}
\label{eq2.12}
\varphi _{n \varepsilon}^{int}({\bf q},{\bf p},{\bf q'},t,t') =
\mbox{Sp} \bigg[ I_n({\bf q},{\bf} p,t) T_q(t,t')
 \int\limits_{0}^{1} d\tau \varrho _q^{\tau}(t')
I_{\varepsilon}^{int}({\bf q'},t') \varrho _q^{1-\tau}(t') \bigg],
\end{equation}
\begin{equation}
\label{eq2.13}
\varphi _{\varepsilon n}^{int}({\bf q},{\bf q'},{\bf p},t,t') =
\mbox{Sp} \bigg[ I_{\varepsilon}^{int}({\bf q},t')  T_q(t,t')
 \int\limits_{0}^{1} d\tau \varrho _q^{\tau}(t') I_n({\bf p},{\bf q'},t)
\varrho _q^{1-\tau}(t') \bigg] ,
\end{equation}
\begin{equation}
\label{eq2.14}
\varphi _{n n}({\bf q},{\bf p},{\bf q'},{\bf p'},t,t') =
\mbox{Sp} \bigg[ I_n({\bf q},{\bf p},t)  T_q(t,t')
 \int\limits_{0}^{1} d\tau \varrho _q^{\tau}(t') I_n({\bf p'},{\bf q'},t)
\varrho _q^{1-\tau}(t') \bigg] ,
\end{equation}
\begin{equation}
\label{eq2.15}
\varphi _{\varepsilon \varepsilon}^{int}({\bf q},{\bf q'},t,t') =
\mbox{Sp} \bigg[I_{\varepsilon}^{int}({\bf q},t') T_q(t,t')
\int\limits_{0}^{1} d\tau \varrho_q^{\tau}(t')I_{\varepsilon}^{int,int}({\bf q'},t')
\varrho _q^{1-\tau}(t')\bigg] .
\end{equation}

The system of equations (\ref{eq2.11}) for the one-particle distribution
function and the average density of potential energy is strongly nonlinear
and it can be used to description both the strongly and weakly
nonequilibrium states of the Bose system with a consistent consideration
of kinetics and hydrodynamics.  The description of weakly nonequilibrium
processes was reviewed in \cite{l49}. Projecting transport equations on
the values of the component of the vector ${\bf\Psi}({\bf{p}})=\left(
{1,\,{\bf p},\,\frac{p^2}{2m} - \frac{q^2}{8m}}\right),$ we shall obtain
the equations of nonlinear hydrodynamics, in which the transport processes
of kinetic and potential parts of energy are described by two
interdependent equations. Obviously, such equations of the nonlinear
hydrodynamic processes give more opportunity to describe the process of
mutual transformation of kinetic and potential energy in detail at
investigation of nonequilibrium processes occurring in the system.

The proposed scheme of transport equations is inconvenient when
considering the kinetics and hydrodynamics in the vicinity of the phase
transition point, where the large-scale fluctuations play the essential
role.

\section{Kinetic equation for the nonequilibrium Wigner  func\-tion
and Fokker-Planck equation for distribution function of the hydrodynamic
variables}

As previously, the nonequilibrium quantum distribution function
$f_1({\bf{q}},{\bf{p}};t)=\langle \hat{n}_{\bf q}({\bf p})\rangle^t$ is
chosen as parameter to description of one-particle correlations. We
introduce the distribution function of hydrodynamic variables to
description of the collective processes in a quantum system as follows:
\begin{equation}
\label{eq3.1}
\hat {f}(a) = \frac{1}{(2\pi)^5}\int d{\bf{x}}\,\mbox{e}^{i{\bf{x}}(\hat{\bf{a}}-\bf{a})},
\end{equation}
where
$\hat{\bf{a}}=\{\hat{a}_{1\bf{k}},\hat{a}_{2\bf{k}},\hat{a}_{3\bf{k}} \}$,
\,  $\hat {a}_{1{\bf k}} = \hat {n}_{\bf k} $,
 $\hat {a}_{2{\bf k}}= \hat {\bf P}_{\bf k}$,
 $\hat {a}_{3{\bf k}} = \hat {\varepsilon}_{\bf k}=\hat {\varepsilon }_{\bf k}^{kin}+\hat {\varepsilon
}_{\bf k}^{int}$
are the Fourier-components of the operators of particles number, momentum
and energy densities (\ref{eq2.1})--(\ref{eq2.4}). The scalar values
$a_{m{\bf{k}}}=\left\{n_{\bf{k}},\,\,{\bf{P}}_{\bf{k}},\,\,
\varepsilon_{\bf{k}}\right\}$ are the corresponding collective variables.
The operator function (\ref{eq3.1}) is obtained in accordance with Weyl
correspondence rule from the classical distribution function \cite{l26}
\begin{equation}
\nonumber
f(a) = \delta({\bf{A}-\bf{a}})=\prod\limits_{m=1}^N \prod\limits_{\bf{k}}
\delta({A}_{m\bf{k}}-a_{m\bf{k}}),
\end{equation}
where ${\bf A}=\{A_{1\bf k}\ldots,A_{N\bf k}\}$ is the classical dynamical
variables.

The average values $f_1({\bf q},{\bf p};t)=\langle \hat {n}_{\bf k}({\bf
p})\rangle^t$, $f(a;t)=\langle \hat{f}(a)\rangle^t$ are calculated using
the nonequilibrium statistical operator $\hat{\varrho}(t)$, which satisfy
the Liouville equation. In line with the idea of abbreviated description
of the nonequilibrium state, the statistical operator $\hat{\varrho}(t)$
must functionally depend on the quantum one-particle distribution function
and distribution functions of the hydrodynamic variables:
\begin{equation}
\label{eq3.2}
\hat {\varrho }(t) =
\hat {\varrho }\left( {\ldots \,f_1 ({\bf q},{\bf p};t),\,f(a;t)\,\ldots } \right).
\end{equation}
Thus, the task is to find the solution of the  Liouville equation for
$\hat{\varrho}(t)$ which has the form (\ref{eq3.2}). For that purpose we
use the method of Zubarev nonequilibrium statistical  operator
\cite{l23,l44,l45,l47}. We consider the Liouville equation  (\ref{eq2.5})
with infinitely small source. The source correctly selects retarded
solutions in accordance with the abbreviated description of nonequilibrium
state of a system. The quasiequilibrium statistical operator
$\hat{\varrho}_q(t)$ is determined in the usual way, from the condition of
the maximum informational entropy functional with the normalization
condition: $\mbox{Sp}\,\hat{\varrho}_q(t)\,=\,1.$ Then the
quasiequilibrium statistical operator can be written as
\begin{equation}
\label{eq3.3}
\hat{\varrho }_q (t) = \exp
\left\{
  -\Phi(t)- \sum\limits_{\bf q} \sum\limits_{\bf p} \gamma _{- {\bf q}}({\bf p};t)
  \hat {n}_{\bf q}({\bf p})- \int {da\,F(a;t)\,\hat {f}(a)}
\right\},
\end{equation}
where $ da\,\, \to \,\,\{dn_{\bf k} ,\,\,\,d{\bf P}_{\bf k}
,\,\,\,d\varepsilon_{\bf k}\}$.

The Massieu-Plank functional $\Phi(t)$ is determined from the
normalization condition:
\[
\Phi (t)= \ln \,\mbox{Sp}\,
\left[ \exp \left\{ - \sum\limits_{{\bf q}{\bf p}} \gamma_{-{\bf q}}({\bf p};t)
\hat {n}_{\bf q}({\bf p}) - \int da\,F(a;t)\,\hat {f}(a) \right\}\right].
\]
Functions $\gamma _{-{\bf q}}({\bf p};t)$ and $F(a,t)$ are the Lagrange
multipliers and can be defined from the self-consistent conditions:
\begin{equation}    \label{eq3.4}
f_1({\bf q},{\bf p};t)= \langle \hat{n}_{\bf q}({\bf p})\rangle^t =
\langle \hat{n}_{\bf q}({\bf p})\rangle_q^t, \quad
f(a;t)= \langle \hat{f}(a)\rangle^t = \langle \hat{f}(a)\rangle_q^t .
\end{equation}
It is convenient to rewrite the quasiequilibrium statistical operator
(\ref{eq3.3}) in the following form:
\begin{equation}
\label{eq3.5} \hat{\varrho }_q (t) =
 \int da \int_{0}^{1}d\tau (\hat{\varrho}_q^{kin}(t))^{\tau}\mbox{e}^{-F(a;t)\hat
 {f}(a)}(\hat{\varrho}_q^{kin}(t))^{1-\tau},
\end{equation}
where
\begin{equation}
\label{eq3.6} \hat{\varrho}_q^{kin}(t) =
\exp\left\{-\Phi(t)-\sum\limits_{{\bf q}\,{\bf p}} \gamma_{-{\bf
q}}({\bf p};t)\hat{n}_{\bf q}({\bf p})\right\}
\end{equation}
is the ''kinetic'' quasiequilibrium statistical operator. Using the
self-consistent conditions (\ref{eq3.4}) we find the function $F(a;t)$:
\begin{equation}
\label{eq3.7}
\mbox{e}^{-F(a;t)} = \int da'\,W_{-1}(a,a';t)f(a';t),
\end{equation}
where function $W_{-1}(a,a';t)$ is determined from integral equation
\begin{equation}
\label{eq3.8}
\int da''\,W(a,a'';t)\,W_{-1}(a'',a';t)=\delta (a-a').
\end{equation}
\begin{equation}
\label{eq3.9}
W(a,a';t) = \mbox{Sp}\left[\hat f(a)\int^1_0 d\tau(\hat\rho_q^{kin}(t))^{\tau}
         \hat f(a')(\hat\rho_q^{kin}(t))^{1-\tau}\right]
\end{equation}
is the structural distribution function of hydrodynamic fluctuations
averaged with "kinetic" quasiequilibrium operator. The functions
$W(a,a';t)$ and $W_{-1}(a,a';t)$ satisfy equation (\ref{eq3.8}) and
contain the singular and regular terms \cite{l26}:
\begin{eqnarray}
\label{eq3.10}
  W(a,a';t)=W(a,t)[\delta(a-a')-R(a,a';t)],
\nonumber   \\
  W_{-1}(a,a';t)=W^{-1}(a,t)[\delta(a-a')-r(a,a';t)],
            \\
  \int da \,W(a;t)\,R(a,a';t)=\int da' \,R(a,a';t)=0,
\nonumber   \\
\int da \,W(a;t)\,r(a,a';t)=\int da' \,r(a,a';t)=0.
\nonumber
\end{eqnarray}

This function can be consider as a Jacobian of the transition in the
collective variables space
$n_{\bf{k}},{\bf{P}}_{\bf{k}},\varepsilon_{\bf{k}}$, which are averaged
with the ''kinetic'' quasiequilibrium statistical operator.

Taking into account (\ref{eq3.7}), the initial quasiequilibrium operator
(\ref{eq3.5}) can be represented as \cite{l26}:
\begin{equation}
\label{eq3.11}
 \hat{\varrho }_q (t) =
 \int da \, \int da'
 \int_{0}^{1}d\tau (\hat{\varrho}_q^{kin}(t))^{\tau}
 \hat{f}(a)(\hat{\varrho}_q^{kin}(t))^{1-\tau}\, W_{-1}(a,a';t)f(a';t),
\end{equation}
or
\begin{equation}
\label{eq3.12}
 \hat{\varrho }_q (t) =
 \int da f(a;t) \hat{\varrho}_{L}(t),
\end{equation}
where
\begin{equation}
\label{eq3.13}
 \hat{\varrho}_L(a;t) =
 \int da' \, \int_{0}^{1}d\tau (\hat{\varrho}_q^{kin}(t))^{\tau}
 \hat{f}(a';t)\, (\hat{\varrho}_q^{kin}(t))^{1-\tau}.
\end{equation}

Gibbs entropy which corresponds to the quasiequilibrium statistical
operator (\ref{eq3.11}) can be written in the form:
\begin{eqnarray}
\label{eq3.14} S(t) =  - \langle \ln \hat{\varrho}_q(t)
\rangle_q^t = \Phi(t) + \sum\limits_{{\bf q}{\bf p}} \gamma_{-{\bf
q}}({\bf
p};t) \langle \hat{n}_{\bf q}({\bf p})\rangle^t_{q} - \\
\int da f(a;t) \ln \left(\int da'W_{-1}(a,a')f(a';t) \right),
\nonumber
\end{eqnarray}
from which, taking into account the self-consistency condition
(\ref{eq3.4}), we obtain nonequilibrium entropy of the Bose-system:
\begin{equation}
\label{eq3.15}
S(t) = \Phi(t) + \sum\limits_{{\bf q}{\bf p}}
\gamma_{-{\bf q}}({\bf p};t) \langle \hat{n}_{\bf q}({\bf
p})\rangle^t - \int da f(a;t) \ln \left(\int
da'W_{-1}(a,a')f(a';t) \right),
\end{equation}
that contains the kinetic and hydrodynamic contributions.

After constructing the quasiequilibrium statistical operator
(\ref{eq3.12}), the Liouville equation (\ref{eq2.5}) for the operator
$\Delta\hat{\varrho}(t)=\hat{\varrho}(t)-\hat{\varrho }_q(t)$ is written
in the form:
\begin{equation}
\label{eq3.16}
\left( \frac{\partial}{\partial t} + i\hat{L}_N + \varepsilon\right)\Delta \hat {\varrho}(t)=
\left( \frac{\partial}{\partial t} + i\hat{L}_N  \right)\hat{\varrho}_q(t).
\end{equation}
Time derivative of the right-hand side of this equation can be expressed
through the projection Kawasaki-Gunton operator  $P_q(t)$
\cite{l23,l26,l47}:
\begin{equation}
\label{eq3.17}
\frac{\partial}{\partial t}\hat{\varrho }_q(t)= - P_q(t)i\hat{L}_N \hat{\varrho}(t).
\end{equation}
In our case the projection operator acts arbitrary on statistical
operators $\hat{\varrho}'$ according to the rule
\begin{eqnarray}
\label{eq3.18}
 P_q(t)\hat{\varrho}'=\hat{\varrho}_q(t)\mbox{Sp}\hat{\varrho}' +
 \sum\limits_{{\bf q}\,{\bf p}}
 \frac{\partial \hat{\varrho}_q(t)}{\partial \langle \hat{n}_{\bf q} ({\bf p}) \rangle^t}
 \left[\mbox{Sp}\left( \hat{n}_{\bf q}({\bf p})\hat{\varrho}' \right) -
 \langle\hat{n}_{\bf q}({\bf p})\rangle^t \mbox{Sp}\hat{\varrho}'\right]
      \\
 + \int{da}\frac{\partial\hat{\varrho }_q(t)}{\partial f(a;t)}
 \left[{\mbox{Sp}\left({\hat{f}(a)\hat{\varrho}'}\right)
 - f(a;t)\mbox{Sp}\hat{\varrho}'}\right]. \qquad \qquad
 \nonumber
 \end{eqnarray}
Taking into account relation (\ref{eq3.17}), we rewrite the equation
(\ref{eq3.16}) as follows:
\begin{equation}
\label{eq3.19}
\left( {\frac{\partial}{\partial t} + \left( {1 - P_q (q)} \right)iL_N +
\varepsilon}\right)\Delta \hat {\varrho }(t) =
-\left( {1 - P_q(q)} \right)i\hat{L}_N\hat{\varrho}_q(t).
\end{equation}
Formal solution of (\ref{eq3.19}) is
\begin{equation}
\label{eq3.20}
\Delta \hat{\varrho(t)} = -\int\limits_{-\infty}^t dt'\,\mbox{e}^{\varepsilon(t' - t)}
T_q (t;t')\left({1-P_q(t)}\right) i\hat{L}_N \hat{\varrho}_q(t),
\end{equation}
where
\begin{equation}
\label{eq3.21}
T_q(t;t') = \exp_{+}
\left\{
-\int\limits_{t'}^t dt' \left({1-P_q (t')} \right)
i\hat{L}_N
\right\}
\end{equation}
is the generalized time evolution operator, that take into account
projection. From (\ref{eq3.20}) we find the nonequilibrium statistical
operator
\begin{equation}
\label{eq3.22}
\hat{\varrho}(t) = \hat {\varrho }_q (t) -
\int\limits_{-\infty}^t dt'\,\mbox{e}^{\varepsilon(t' - t)}
T_q (t;t')\left({1-P_q(t)}\right) iL_N \hat{\varrho }_q (t).
\end{equation}
Now we consider the operation of  Liouville  operator on the
quasiequilibrium operator (\ref{eq3.11}):
\begin{eqnarray}
\label{eq3.23}
 i\hat{L}_N \hat{\varrho }_q (t)=
 -\sum\limits_{{\bf q}\,{\bf p}} {\gamma_{-{\bf q}} }({\bf p};t)
 \int\limits_{0}^{1} d\tau (\hat{\varrho}_q(t))^{\tau}
 \dot{\hat{n}}_{\bf q}({\bf p})(\hat{\varrho}_q(t))^{1-\tau}
\\
-\int da F(a;t) \int\limits_{0}^{1} d\tau(\hat{\varrho}_q(t))^{\tau}
i\hat{L}_N \hat{f}(a)(\hat{\varrho}_q(t))^{1-\tau},
\nonumber
\end{eqnarray}
where
$\dot{\hat{n}}_{\bf q}({\bf p}) = i\hat{L}_N \hat{n}_{\bf q}({\bf p})$.
Since \cite{l26}
\begin{equation}  \label{eq3.24}
i\hat L_N \hat f(a)=-\frac{\partial}{\partial a}\hat J(a)
\end{equation}
with
\begin{equation}
\label{eq3.25}
\hat J(a)=(2\pi)^{-N}\int dx e^{ix(\hat{a}-a)}
\int_0^1 d\tau e^{-i\tau x\hat{a}} i\hat{L}_N \hat{a}e^{i\tau x\hat{a}},
\end{equation}
the second term on the right-hand side of (\ref{eq3.23}) can be
represented:
\begin{eqnarray}
\int da F(a;t) \int\limits_{0}^{1}d\tau (\hat{\varrho}_q(t))^{\tau}
(-\frac{\partial}{\partial a}\hat{J}(a)) (\hat{\varrho}_q(t))^{1-\tau}=
\nonumber \\
\int da (\frac{\partial}{\partial a}F(a;t)) \int\limits_{0}^{1}d\tau
(\hat{\varrho}_q(t))^{\tau} \hat{J}(a)(\hat{\varrho}_q(t))^{1-\tau}
\nonumber
\end{eqnarray}
and using (\ref{eq3.7}) it can be written as follows:
\begin{eqnarray}
\label{eq3.26}
\int da (\frac{\partial}{\partial a}F(a;t)) \int\limits_{0}^{1}d\tau
(\hat{\varrho}_q(t))^{\tau} \hat{J}(a)(\hat{\varrho}_q(t))^{1-\tau}=
                         \qquad \qquad
\\
-\int da \left[
\frac{\partial}{\partial a} \ln \int da' W_{-1}(a,a')f(a';t)
\right]
\int\limits_{0}^{1} d\tau (\hat{\varrho}_q(t))^{\tau}\hat J(a)
(\hat{\varrho}_q(t))^{1-\tau}.
\nonumber
\end{eqnarray}
Then the expression (\ref{eq3.23}) taking into account (\ref{eq3.23})
will take the form:
\begin{eqnarray}
\label{eq3.27} i\hat{L}_N \hat\varrho_q(t)=-\sum\limits_{{\bf
q},{\bf p}}{\gamma_{-{\bf q}} }({\bf p};t) \dot{\hat{n}}_{\bf
q}({\bf p};\tau) \hat{\varrho}_{q}(t)+
\\
\int da \hat{J}(a;\tau)\left[
\frac{\partial}{\partial a} \ln \int da' W_{-1}(a,a';t)f(a';t)
\right]   \hat{\varrho}_q(t),
\nonumber
\end{eqnarray}
where
\begin{equation}
\label{eq3.28}
\dot{\hat{n}}_{\bf q}({\bf p};\tau)=\int\limits_{0}^{1} d\tau
(\hat{\varrho}_q(t))^{\tau}\dot{\hat{n}}_{\bf q}({\bf p})
(\hat{\varrho}_q(t))^{-\tau},
\hat{J}(a,\tau)=\int\limits_{0}^{1}d\tau
(\hat{\varrho}_q(t))^{\tau}\hat{J}(a)
(\hat{\varrho}_q(t))^{-\tau}.
\end{equation}

Now in accordance with (\ref{eq3.26}) and (\ref{eq3.12}) the Liouville
operator action on $\hat{\varrho}_q(t)$ can be represented as:
\begin{eqnarray}
\nonumber
i\hat{L}_N \hat{\varrho}_q(t) =-\int da \sum\limits_{{\bf q},{\bf p}}
\gamma_{-{\bf q}}({\bf p};t)\dot{\hat{n}}_{\bf q}({\bf p},\tau) f(a;t)
\hat{\varrho}_L(a;t)
\nonumber   \\
+ \int da \int a'' \sum\limits_{\bf k}
\left[
 \hat{J}(n_{\bf k};\tau)\frac{\partial}{\partial n_{\bf k}}+
 \hat{J}({\bf P}_{\bf k};\tau)
  \frac{\partial }{\partial {\bf P}_{\bf k}}+
 \hat{J}({\varepsilon}_{\bf k};\tau)
\frac{\partial}{\partial \hat{\varepsilon}_{\bf k}(\tau)} \right]
\nonumber \\
\times
\ln\left[\int da' W_{-1}(a,a';t) f(a';t) \right] f(a'';t)
\hat{\varrho}_L(a'';t').  \qquad \qquad
\nonumber
\end{eqnarray}
Finally, substituting this expression in (\ref{eq3.22}), we obtain the
nonequilibrium statistical operator:
\begin{eqnarray}
\label{eq3.29}
\hat{\varrho}(t) = \int da f(a;t) \hat{\varrho}_L(a,t) + \int da
\sum\limits_{{\bf q}\,{\bf p}} \int\limits_{-\infty }^t dt'
\mbox{e}^{\varepsilon (t' - t)}T_q (t,t')
   \\
\times \left({1 - P_q(t')}\right) \dot{\hat{n}}_{\bf q}({\bf
p},\tau)\hat{\varrho}_L(a;t')f(a;t)\gamma_{-{\bf q}}({\bf p};t')
\nonumber   \\
+\int da \int a'' \sum\limits_{\bf k} \int\limits_{-\infty}^t dt'
 \mbox{e}^{\varepsilon (t' - t)}T_q (t,t')
 \left( {1 - P_q(t')} \right)
\nonumber   \\
\times
\left[
 \hat{J}(n_{\bf k};\tau)\frac{\partial}{\partial n_{\bf k}}+
 \hat{J}({\bf P}_{\bf k};\tau)
  \frac{\partial }{\partial {\bf P}_{\bf k}}+
 \hat{J}({\varepsilon}_{\bf k};\tau)
\frac{\partial}{\partial \hat{\varepsilon}_{\bf k}(\tau)}
\right]
\nonumber \\
 \times \ln\left[\int da' W_{-1}(a,a';t) f(a';t) \right] f(a'';t)
\hat{\varrho}_L(a'';t').
\nonumber
\end{eqnarray}

This  formula gives the nonequilibrium statistical operator that
consistently  describes the kinetic and nonlinear hydrodynamic
fluctuations of quantum Bose fluids. After neglecting the hydrodynamic
fluctuations we can return to the traditional scheme accepted in the
kinetic theory. Nonequilibrium statistical operator is the functional of
the abbreviated description parameters  $f_1(\bf{q, p}; t)$, $f(a;t)$,
which are required for complete description of the transport equations.
For this we use relations:
\begin{eqnarray}
\label{eq3.30}
 \frac{\partial}{\partial t} f_1({\bf q},{\bf p};t) =
 \langle \dot{\hat {n}}_{\bf q}({\bf p})\rangle^t =
 \langle \dot{\hat {n}}_{\bf q} ({\bf p})\rangle_q^t
 + \langle I_n ({\bf q},{\bf p})\rangle^t,
 \nonumber    \\
 \frac{\partial}{\partial t}f(a;t) = \mbox{Sp}
 \left\{\hat{\varrho}_L(a,t)i\hat{L}_N \hat{f}(a)\right\},
 \end{eqnarray}
where $I_n({\bf q},{\bf p};t)$ is the generalized flow of density. Thus,
substitution of explicit expression (\ref{eq3.29}) into these relations
and after simple but somewhat unwieldy transformations, we obtain the
final expressions for kinetic equations:
\begin{eqnarray}
\label{eq3.31}
 \left( \frac{\partial}{\partial t}-i\frac{{\bf q}{\bf p}}{m} \right)
 f_1({\bf q},{\bf p};t) - \int da f(a;t')\langle i\hat{L}_N^{int}\hat {n}_{\bf q}({\bf p})\rangle_L^t
                 \\
= \sum\limits_{{\bf q}'\,{\bf p}'} \int da \int\limits_{-\infty}^t dt'
 \mbox{e}^{\varepsilon (t' - t)}
 \Phi_{nn}({\bf q},{\bf q}',{\bf p},{\bf p}';t,t')f(a;t')\gamma _{{\bf q}'}({\bf p}';t)
 \nonumber       \\
 - \sum\limits_{{\bf q}'\,{\bf p}'} \sum\limits_{\bf k}
 \int da \int da'' \int\limits_{-\infty }^t dt'
 \mbox{e}^{\varepsilon(t' - t)}
\left\{
 \Phi_{n{\bf P}}({\bf q},{\bf p},a,a'';t,t')
 \frac{\partial}{\partial{\bf P}_{\bf k}}
\right.
 \nonumber   \\
\left.
+ \Phi_{n \varepsilon}({\bf q},{\bf p},a,a'';t,t')
 \frac{\partial }{\partial \varepsilon_{\bf k} }
\right\}
 \left[\ln\int da' W_{-1}(a,a';t)f(a,';t)\right]f(a'';t),
\nonumber
\end{eqnarray}
\begin{eqnarray}
\label{eq3.32}
 \frac{\partial}{\partial t}f(a;t) + \sum\limits_{\bf k}
\left\{
   \frac{\partial}{\partial n_{\bf k}} \int da'v_{n}(a,a';t)f(a';t)
\right.
\nonumber      \\
\left.
 + \frac{\partial}{\partial {\bf P}_{\bf k}} \int da' v_{\bf P}(a,a';t)f(a';t)
 + \frac{\partial}{\partial \varepsilon_{\bf k}}
    \int da' v_{\varepsilon}(a,a';t)f(a';t)
\right\}       \\
\nonumber
 = \sum\limits_{\bf k} \frac{\partial}{\partial {\bf
P}_{\bf k}}
  \sum\limits_{{\bf q}'\,{\bf p}'} \int da'  \int\limits_{-\infty}^t dt'
  \mbox{e}^{\varepsilon(t'-t)}
 \times
  \Phi_{{\bf P}n}(a,a'',{\bf q}',{\bf p}';t,t')f(a';t')
  \gamma_{{\bf q}'}({\bf p}';t')
  \nonumber     \\
+ \sum\limits_{\bf k} \frac{\partial}{\partial \varepsilon_{\bf k}}
 \sum\limits_{{\bf q}'\,{\bf p}'} \int da'  \int\limits_{-\infty}^t dt'
 \mbox{e}^{\varepsilon (t'-t)}
 \Phi_{\varepsilon n}(a,a',{\bf q}',{\bf p}';t,t')f(a';t')
 \gamma_{{\bf q}'}({\bf p}';t')
 \nonumber
 \end{eqnarray}
\begin{eqnarray}
- \sum\limits_{{\bf k}\,{\bf q}'} \int da'' \int\limits_{-\infty}^t dt'
 \mbox{e}^{\varepsilon (t'-t)}
\left[\frac{\partial}{\partial {\bf P}_{\bf k} }
 \Phi_{{\bf PP}}(a,a',a'';t,t')
 \frac{\partial}{\partial{\bf P}_{\bf q}}
\right.
 \nonumber      \\
\left.
 +\frac{\partial}{\partial {\bf P}_{\bf k} }
 \Phi_{{\bf P}\varepsilon}(a,a',a'';t,t')
 \frac{\partial}{\partial \varepsilon_{\bf q}}
 +\frac{\partial}{\partial \varepsilon_{\bf k} }
  \Phi_{\varepsilon {\bf P}}(a,a',a'';t,t')
  \frac{\partial}{\partial {\bf P}_{\bf q}}
\right.
 \nonumber      \\
\left.
+ \frac{\partial}{\partial \varepsilon_{\bf k}}
  \Phi_{\varepsilon \varepsilon}(a,a',a'';t,t')
  \frac{\partial}{\partial \varepsilon_{\bf q}}
\right]
   \left[\ln \int da' W_{-1}(a,a';t)f(a';t)\right]
  f(a'';t),
\nonumber
\end{eqnarray}
where $i\hat{L}_N^{int}$ is a potential part of the Liouville operator,
\begin{equation}
\label{eq3.33}
 v(a,a';t)=\langle \hat{J}(a,a')\rangle_L^t = \int da''
 \mbox{Sp}(\hat{J}(a)\int\limits_{0}^{1} d\tau (\varrho_{q}^{kin})^{\tau}
 f(a'')(\varrho_{q}^{kin})^{1-\tau}W_{-1}(a'',a';t)
\end{equation}
are the generalized flows in collective variables space. Dissipative
processes in the transport equations (\ref{eq3.31}), (\ref{eq3.32}) are
described by the generalized transport kernels (memory functions):
\begin{equation}
\label{eq3.34}
\Phi_{nn}({\bf{q,q',p,p'}},a;t,t') =
\langle I_n({\bf{q,p}};t)T_q(t,t')I_n({\bf{q,p'}};t',\tau)\rangle_L^{t'},
\end{equation}
\begin{eqnarray}
\label{eq3.35}
\Phi_{n\bf P}({{\bf{q,p,k}}},a;t,t') =
 \langle I_n({\bf{q,p}};t)T_q(t,t')I_{\bf P}({\bf{k}};t',\tau)\rangle_L^{t'},
\\
\Phi_{n\varepsilon}({{\bf{q,p,k}}},a;t,t') =
 \langle I_n({\bf{q,p}};t)T_q(t,t')I_{\varepsilon}({\bf{k}};t',\tau)\rangle_L^{t'},
\nonumber
\end{eqnarray}
\begin{eqnarray}
\label{eq3.36}
\Phi_{\bf P\bf P}({{\bf k}},a;t,t') =
 \langle I_{\bf P}({\bf k}; t)T_q(t,t')I_{\bf P}({\bf{k}};t',\tau)\rangle_L^{t'},
\\
\Phi_{\bf P\varepsilon}({\bf{k,q}},a;t,t') =
 \langle I_{\bf P}({\bf k};t)T_q(t,t')I_{\varepsilon}({\bf q};t',\tau)\rangle_L^{t'},
\nonumber
\end{eqnarray}
\begin{equation}           \label{eq3.37}
\Phi_{\varepsilon\varepsilon}({\bf{k,q}},a;t,t') =
 \langle I_{\varepsilon}({\bf k};t)T_q(t,t')I_{\varepsilon}({\bf q};t',\tau)\rangle_L^{t'}.
\end{equation}
In particular, the kernel (\ref{eq3.34}) describes the kinetic processes,
kernels (\ref{eq3.35}) describe the mutual dynamic correlatios between
kinetic and hydrodynamic fluctuations, and (\ref{eq3.36}) describe the
dynamic correlation between the viscous and thermal hydrodynamic
fluctuations. In expressions (\ref{eq3.34})--(\ref{eq3.37}) the
generalized flows are introduced as follows:
\begin{eqnarray}      \label{eq3.38}
I_n({\bf{q,p}};t) = \Big(1-P(t)\Big)\hat{J}(n_{\bf k}), \nonumber
\\
I_{\bf P}({\bf{k}};t,\tau) = \Big(1-P(t)\Big)\hat{J}({\bf P}_{\bf
k}),
\\
I_{\varepsilon}({\bf{k}};t',\tau) =
\Big(1-P(t)\Big)\hat{J}(\varepsilon_{\bf k}), \nonumber
\end{eqnarray}
where $P(t)$ is the Mori projection operator connected to the projection
Kawasaki-Gunton operator (\ref{eq3.18}) $P_q(t)a_{\bf k}\rho_q(t) =
\rho_q(t)P(t)a_{\bf k}$.  It is important to note, that the average
quantities in (\ref{eq3.33})--(\ref{eq3.37}) are calculated using the
quasiequilibrium statistical operator $\rho_L(a; t)$ (\ref{eq3.13}),
therefore the kernels are the functions of collective variables
$a_{\bf{k}}$.

The obtained system of equations (\ref{eq3.31}), (\ref{eq3.32}) takes into
account the effects of time delay and hard nonlocality for variables
$\hat{a}$ associated with the contribution of rapid small-scale
fluctuations and noncommutative base set $\hat{n}_{\bf q}(\bf p)$,
$\hat{a}_{\bf k}= \{ \hat{n}_{\bf k}, {\bf P}_{\bf
k},\hat{\varepsilon}_{\bf k} \}$. Special difficulty is associated with
the structure function $W(a,a';t)$ and averaged velocities $v(a,a';t)$,
containing the singular and regular parts:
\begin{equation}
\label{eq3.39}
 v(a,a';t)=v(a;t)\delta(a-a')+\delta v(a,a';t).
\end{equation}
If taking into account only the singular parts of functions, we proceed to
the local approximation for the statistical operator $\hat{\varrho}(t)$
and the generalized transport equations (\ref{eq3.34}),(\ref{eq3.35}). The
local approximation of generalized Fokker-Planck equation for quantum
systems in general case was discussed in paper \cite{l26} in detail.

\section{Conclusion}

We considered two possible variants for consistent description of
nonlinear kinetic and hydrodynamic processes for quantum Bose systems,
using the Zubarev nonequilibrium statistical  operator.

In the first variant,  the non-equilibrium Wigner distribution function
and the averaged potential energy of interaction were selected as
parameters of abbreviated description. Generalized transport equations for
them are not closed. In order to close them, the spatial gradients of
Lagrange multipliers $\beta_{-{\bf q}}(t)$, $\gamma_{-{\bf q}}({\bf p};t)$
should be determined  from the corresponding self-consistent conditions.

Such calculations, in linear respective to spatial gradients
approximation, are complicated and leads to linearized equations of the
consistent description of kinetics and hydrodynamics of quantum Bose
system \cite{l49}. In higher approximations the problems of solving the
corresponding nonlinear integral equations would emerge.

In our approach, a part of the average interaction potential energy is
connected with the nonequilibrium distribution function of a pair
condensate and can be readily extracted. Further, we can obtain the the
system of equations for time correlation functions with separation of
one-partial and pair-partial Bose condensate both in the Hamiltonian
\cite{l56,l57,l58} and in expressions. Moreover, in our approach we can
start from the quasiequilibrium statistical operator (\ref{eq2.6}) to
construct a chain of BBGKY equations for nonequilibrium particles
distribution functions with the modified boundary conditions (taking into
account multipartial correlations), that is similar both for classical
\cite{l43} and Fermi systems \cite{l58}. In the second variant, the
non-equilibrium Wigner distribution function and nonequilibrium
distribution function of hydrodynamic collective variables (the number
density of particles, their momentum and full energy) were selected as
abbreviated  description parameters. As a result, the system of equations
is obtained. The first equation is of the type of kinetic equation for
nonequilibrium Wigner distribution function with kinetic transport
kernels. The second one is the generalized Fokker-Planck equation for the
nonequilibrium distribution function of collective variables taking into
account nonmarkov effects. In these equations, the averaging procedure is
performed using the quasiequilibrium statistical operator (\ref{eq3.15}),
in which the structural function $W(a,a';t)$ (it contains singular and
regular parts) is an average of the transition operator $\hat f(a)$ from
phase variables to collective variables $a$. Calculation of $W(a,a';t)$ is
a key moment here. This is because this function enters the Fokker-Planck
equation and is used for calculation of the hydrodynamic speeds
$v_{l}(a,a';t)$ (it contains singular and regular parts) and generalized
kernels (\ref{eq3.34})--(\ref{eq3.37}). In the case studies of nonlinear
hydrodynamic fluctuations \cite{l59,l60} or the consistent description of
kinetic and nonlinear hydrodynamic fluctuations \cite{l61,l62} for
classical systems, the structural function $W(a;t)$ was calculated by the
collective variables method \cite{l63}. It opens the opportunity to go
beyond the Gaussian approximation for $W(a;t)$ and to propose an approach
to calculation of the generalized transport coefficients in higher
approximations for the fluctuations, in particular, of the coupled modes
type \cite{l60,l64}. Calculation of the structure function $W(a,a';t)$ and
$v_{l}(a,a';t)$ in local approximations for quantum Bose systems requires
an individual consideration and will be presented in the next paper.

\end{document}